\documentclass[10pt,journal,compsoc]{IEEEtran}

\ifCLASSOPTIONcompsoc
  \usepackage[nocompress]{cite}
  \usepackage{algorithm}
  \usepackage{multirow}
  \usepackage{graphicx}
  \usepackage{algorithm}
  \usepackage{algorithmic}
  \usepackage{array}
  \usepackage{subfigure}
  \usepackage{amsthm}

  \usepackage[justification=centering]{caption}
  \usepackage{url}
  \usepackage{cite}
  \usepackage{setspace}
  \usepackage{amsmath,amssymb,amsfonts}
  \usepackage{algorithmic}
  \usepackage{graphicx}
  \usepackage{textcomp}
  \usepackage{enumerate}
  \usepackage{amsfonts}
  \usepackage{amsmath} 
  \usepackage{amssymb}
  \usepackage{verbatim}
  \usepackage{float}
  \usepackage{setspace}
  \usepackage{threeparttable}
\else
  \usepackage{cite}
  
\fi

\ifCLASSINFOpdf

\else

\fi

\begin{document}

\title{Optimal Data Placement for Data-Sharing Scientific Workflows in Heterogeneous Edge-Cloud Computing Environments}
%
%
%
%

\author{Xin Du,~\IEEEmembership{Student Member,~IEEE,}
	Songtao Tang,~\IEEEmembership{}   
	Zhihui Lu,~\IEEEmembership{Member,~IEEE,}
	Keke Gai,~\IEEEmembership{Senior Member,~IEEE,}
	Jie Wu,~\IEEEmembership{Member,~IEEE}
	and~Patrick C.K. Hung,~\IEEEmembership{Senior Member,~IEEE,}
	\IEEEcompsocitemizethanks{\IEEEcompsocthanksitem The previous version of this
		work titled “A Novel Data Placement Strategy for Data-Sharing Scientific Workflows in Heterogeneous Edge-Cloud Computing Environments” was
		published in 2020 IEEE International Conference on Web Services (ICWS), Beijing, China, 2020, pp. 498-507~\cite{9284088}. (Corresponding authors: Zhihui Lu.)
	\IEEEcompsocthanksitem This version is intended only to provide an idea for others to continue exploring and quoting the published version. The specific contents of this manuscript need to be fully supplemented and adjusted.
		}
	\thanks{Manuscript received XXX; revised XXX.}

}
%
%

\markboth{Journal of \LaTeX\ Class Files,~Vol.~14, No.~8, August~2015}%
{Shell \MakeLowercase{\textit{et al.}}: Bare Demo of IEEEtran.cls for Computer Society Journals}
%



\IEEEtitleabstractindextext{%
\begin{abstract}
The heterogeneous edge-cloud computing paradigm can provide a more optimal direction to deploy scientific workflows than traditional distributed computing or cloud computing environments. Due to the different sizes of scientific datasets and some of these datasets must keep private,  it is still a  difficult problem to finding an data placement strategy that can minimize data transmission as well as  placement cost. To address this issue, this paper combines advantages of both edge and cloud computing to construct a data placement model, which can balance data transfer time and data placement cost using intelligent computation. The most difficult research challenge the model solved is to consider many constrain in this hybrid computing environments, which including shared datasets within individual and among multiple workflows across various geographical regions. According to the constructed model, the study propose a new data placement strategy named DE-DPSO-DPS, which using a discrete particle swarm optimization algorithm with differential evolution (DE-DPSO-DPA) to distribute these scientific datasets.  The strategy also not only consider the characteristics such as the number and storage capacity of edge micro-datacenters, the bandwidth between different datacenters and the proportion of private datasets, but also analysis the performance of algorithm during the workflows execution. Comprehensive experiments are designed in simulated heterogeneous edge-cloud computing environments demonstrate that the data placement strategy can effectively reduce the data transmission time and placement cost as compared to traditional strategies for data-sharing scientific workflows.
\end{abstract}

}

\maketitle

\IEEEdisplaynontitleabstractindextext

\IEEEpeerreviewmaketitle

\IEEEraisesectionheading{\section{Introduction}\label{sec:introduction}}
\IEEEPARstart{R}{ecent} years, the exponential increase of global cooperation in the scientific research and the rapid development of distributed computing technology have resulted in significantly change of scientific applications. They now involve thousands of interwoven tasks and are generally data and computing intensive~\cite{kashlev2015typetheoretic}. Because of this, scientific workflow is widely used to represent these complicated applications in several scientific fieldss~\cite{li2015composition}. Such as these field of astronomy, physics, and bioinformatics, their datasets generally have complex structure and different sizes, so the deployment of their scientific workflows has rigid requirements for computational and storage resources. Specially, in traditional distributed and cloud computing environments, the data transmission delay during the execution of scientific workflows is not satisfactory for the requirement of scientific cooperation. Furthermore, the maintainance and the construction of these traditional scientific workflow systems are very expensive. To achieve this challenge, the worthy data placement model and strategy must be consider in complex practical workflows' scenarios. The datasets are often shared among multiple tasks within workflows, including workflows in different geo-distributed organizations. Moreover, there are a lot of private datasets that may only be allowed to be stored in specific research institutes.

The emergence of heterogeneous edge-cloud computing provide a more optimal paradigm to satisfy the demand of the data placement strategy for scientific workflows. Cloud computing has supported with large-scale commodity hardware, which virtualizes infinite resources with lower construction and maintenance cost. When deploying scientific workflow, it is high efficiency, flexibility, scalability, cost-efficient but the remote end lead to serious transmission delay and private security problem~\cite{duc2019machine,wu2019qamec,lu2018iotdem}. Edge computing can reduce the data transmission delays and guarantee the copyright and privacy of the scientific datasets. Unfortunately, the limited resources and more expensive cost result in the phenomenon that edge computing resources can't store all the datasets~\cite{zhang2018efficient}. The heterogeneous edge-cloud computing environment~\cite{wu2019mobility,shi2017edge,xiao2020orhrc}, which ensures the resource supply and the security of private datasets, combine the advantages of both edge computing and cloud computing to make optimal tradeoff of minimizing the data transmission time and the placement cost. As shown in Figure 1, the environment covers a lot of datacenters, including cloud datacenters that are distributed geographically and edge micro-datacenters that are in the near end users. 

Different from the cloud environment, which are already several successful scientific cloud workflow systems, the applyment of heterogeneous edge-cloud computing environments have to systematically tackle a lot of challenges. As mentioned in this study, one of key issues is the data placement problem.  According to the real-world scenarios, a practicable data placement strategy should satisfy these conditions: Firstly, scientific workflows which needed to support the scientific collaboration between different research institutes, has the characteristics of distributed and data-intensive. During the process of workflow execution, datasets sharing among multiple workflows and tasks usually be used to improve the efficiency of the whole systems. Furthermore, these scientific datasets and tasks sometimes will be allocated and dispatched between geographically distributed datacenters to facilitate collaborative different research. Secondly, due to the large number of datasets and complex structures of scientific workflows, combining edge and cloud computations ensures high cohesion within a datacenter and low coupling between different datacenters. Thirdly, due to the special features of confidentiality and copyright protection of some scientific datasets, its cannot be shared by allocating and dispatching. In other word, some geographically distributed research institutes may own their own private datasets, which only stored in their own micro-edgecenters. Furthermore, there are significant differences in bandwidth and placement cost between different edge micro-datacenters in different geographic locations. These differences have a significant impact on the overall data placement strategy. In summary, the reasonably complement of edge computing and cloud computing can be used to optimize the data placement for data-sharing scientific workflows, but they need an effective mechanism to cooperate for the optimal data placement.

\begin{figure} [!t]
	\setlength{\abovecaptionskip}{0.cm}
	\setlength{\belowcaptionskip}{-0.5cm}

	\caption{Different environment between data placement models. (a) existing heterogeneous edge-cloud computing environment; (b) proposed  heterogeneous edge-cloud computing environment.} 
	\label{fig:mini:subfig} 
\end{figure}

The formulation of the optimal data placement is a NP-hard problem with constraints, such as the storage of these datacenters and placement cost of datasets for different datacenters and so on. 
Contemporarily, Several studies have mapped it to the knapsack packing problem~\cite{du2019service,shao2019data,chen2019idisc,du2019oprc}, and hence, to obtain the optimal solution to this problem, they proposed several methods using heuristic algorithms, such as the genetic algorithm (GA) and particle swarm optimization (PSO). These evolutionary computationary algorthms, which simulate the behaviors of birds, nuts, or fishes in continuous search spaces, are well-suited for solving NP-hard problems. In \cite{li2016novel}, in order to apply the PSO to cloud computing scenarios, a novel discrete PSO (DPSO) algorithm is presented to solve discrete problems. Furthermore, a self-adaptive discrete PSO algorithm with genetic algorithm operators(GA-DPSO) is proposed in~\cite{lin2019time} to reduce the data transmission time during a workflow execution in a single-region heterogeneous edge-cloud computing environment. 

However, these data placement strategies only consider individual workflows or cloud environments, which consider each workflow in isolation. When finding data placement maps in real-world scenarios, some datasets are shared not only among multiple workflows but also among different geographically distributed scientific research institutions. From this data-sharing perspective, figure 2 shows the different environment between the proposed model and others. Unlike the existing heterogeneous edge-cloud computing environement used during the execution of scientific workflows, the showed environement in Figure 2(b) not only considers shared datasets both within tasks and among multiple tasks, but it also considers the coordination of data centers between different regions. To the best of our knowledge, when proposing their data placement strategy, other existing models are not suitable for the actual heterogeneous environment. Meanwhile, due to the complex and so many constrain of heterogeneous edge-cloud environments, this popular research achievements which including our original version mainly focus on single criterion, such as data placement cost or data transfer time. Considering the data placement for scientific workflows in real-world scenarios, a worthy strategy should be based on a trade-off between these two metrics to obtaining the better solutions. 

A data placement strategy proposed in this paper brings a novel data placement model for data-sharing scientific workflows.  According to the characteristics of different datacenters and data-sharing scientific workflows, this paper also presents a discrete PSO algorithm with differential evolution to reduce the data transfer time and data placement cost. During the execution of multiple workflows in multi-region heterogeneous edge-cloud computing environments, the algorithm not only improve the performance of whole system, but also consider many factors impacting the data placement cost and transfer delay. To be specific, we verify the performance of the method in the system by the number of iterations of the algorithm. Meanwhile, the number of datacenters, the storage capacity of edge micro-datacenters, the data sharing for multiple workflows in different datacenters, the bandwidth between different datacenters, the ratio of shared datasets, the number of workflows and the ratio of private datasets are also considered in our data placement model and strategy. we conducted several comprehensive experiments to show that our strategy can effectively reduce data transfer time and placement cost. In summary, our main contributions are as follows.

\begin{enumerate}[1)]  
	\item We model the distributed real-world placement scenarios for optimizing data placement during the execution of multiple workflows in multi- heterogeneous edge-cloud computing environments. Different from previous data placement model, the model does not only considers shared datasets both within tasks and among multiple workflows, but also considers the coordination of data centers between different regions. 
	\item We formulate the data placement problem and propose a novel data placement algorithm. According to the above-mentioned data placement model, the algorithm is based on effective improvement in DPSO and DE to distribute datasets. We recode and define the crossover and mutation operator of the DE to be better suited to our data placement strategy. This algorithm is highly efficient and  has lower complexity to place these datasets in the heterogeneous edge-cloud computing environments.
	\item We propose a data-share data placement strategy based on the data placement model and algorithm for scientific workflows. This strategy considers almost all impact factors that may affect the final results, such as the number of edge micro-datacenters, the storage capacity of edge micro-datacenters, the data sharing for multiple workflows, and the bandwidth between different datacenters.
	\item We evaluate the placement strategy using the data transfer time and placement cost, which are two important indexes by tradeoff to obtaining the better data placement solutions. Moreover, we evaluate our DE-DPSO-DPA by calulating the number of iterations during the process for algorithm gradually converges, which can denote the speed of finding optimal data placement.  
\end{enumerate}

The rest of this study is organized as follows. Section 2 demonstrates the motivation scenario and problem analysis. In section 3, we introduce details of our data placement model and data-sharing workflow notation are defined. Section 4 decribes our data placement strategy.  Section 5 describes and analysis the experimental results. In Section 6, we reviews related work. And the conclusions and future work are summarized in Section 7.

\section{MOTIVATION SCENARIOS AND PROBLEM ANALYSIS}

In this section, we first model the  heterogeneous edge-cloud computing environment, describe the data-sharing scientific workflows execution process. Then, we illustrate a motivating example and analyze the above-mentioned data placement model.
\subsection{The heterogeneous edge-cloud computing environment}
The heterogeneous edge-cloud computing environment $DC = \{DC_{c},DC_{e}\}$ contains geographically distributed cloud datacenters and multiple edge micro-datacenters: cloud computing at the long-distance end, which generally has unlimited storage resources and only one per region, and edge computing in the near end, which is used to store the private datasets with limited storage capacity. According to the heterogeneous computing environment, our model consists of $m$ cloud datacenters $DC_{c} = \{ dc_{1}, dc_{2},.., dc_{m}\}$ and $n$ edge micro-datacenters $DC_{e} = \{ dc_{1}, dc_{2},.., dc_{n}\}$. In addition, $dc_{i} = < cap_{i}, type_{i}>$ represent the $i$th datacenter, whose $cap_{i}$ denotes its storage capacity and $type_{i}$ is used to flag whether the datacenter is a cloud or an edge micro-datacenter. In particular, when $type_{i} = 0$, the datacenter is a cloud  datacenter, which can only store public datasets and is usually distributed geographically.  When $type_{i} = 1$, the datacenter is an edge micro-datacenter, which can store both private and public datasets. The bandwidth across different datacenters is represented as $b_{ij}=<band_{ij},type_{i},type_{j}>$, where $band_{ij}$ is the value of the bandwidth and datacenter $i$ is not datacenter $j$. To improve the focus on our data placement issues, it was assumed that the bandwidth is a known constant. Then, fundamental definitions of data-sharing scientific workflows were given, and the model was constructed.

\subsection{A Working Example}
\begin{figure}[!t]
	\centering
	\includegraphics[width=8 cm]{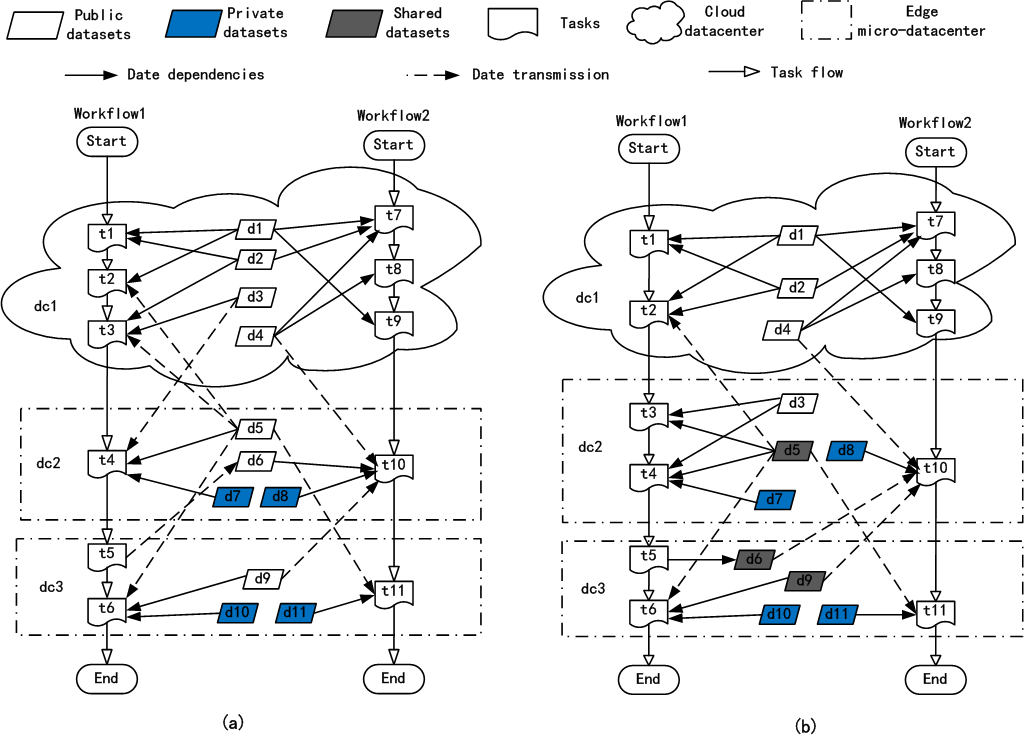}
	\caption{Sample of data placement by two strategies.}
\end{figure} 

In this subsection, a small-scale working example for data-sharing scientific workflows in above-mentioned environment is described to illustrate our data placement strategy. On one hand, as shown in Figure 3(a) and Figure 3(b), there are  two individual workflows in the same region, named workflow 1 and workflow 2. In addition, for the pulsar searching in astrophysic, this simple scientific workflow scenario has 11 tasks $\{t_{1},t_{2},...,t_{11}\}$,  11 datasets $\{d_{1},d_{2},...,d_{11}\}$, and three datacenters $\{dc_{1},dc_{2},dc_{3}\}$. Specifically, four private datasets, which only be deployed in edge micro-datacenters datacenters, are deployed separately in two edge micro-datacenters, $dc_{2}$ and $dc_{3}$, and several datasets are shared between different workflows. It is well known that data transfer speed between cloud datacenters with unlimited storage capacity are lower than the speed between the cloud and edge micro-datacenters. Meanwhile, the bandwidth between the cloud and edge micro-datacenters is much lower than the bandwidth between the edge micro-datacenters~\cite{meng2019dedas}. Corresponding to the data-sharing scientific workflows in Fig 3(a) and Fig 3(b), according to the real datasets in astronomy, table 1 lists the respective sizes of all the datasets in the same-region heterogeneous edge-cloud computing environments. It is worth noting that, due to the sizes of worlflow tasks are much less than scientific datasets, so we ignore them in this paper. 

On the other hand, Figure 4 show that geographically distributed scientific institutes respectively have their own private datasets, workflow systems and data-sharing processes. There are four workflows, two cloud datacenters, and several edge micro-datacenters. Furthermore, there are a lot of datasets and tasks, all of them are scheduled and distributed to these datacenters by data placement strategy. Based on the real scenario, which involved the execution of multiple workflows in multi-region heterogeneous edge-cloud computing environments, the model in this study considers multiple cloud datacenters and multiple scientific workflows, while proposing a data placement strategy based on DE-DPSO, which places datasets while considering not only the bandwidth between datacenters, the number of edge micro-datacenters, and the storage capacity of edge micro-datacenters but also the sharing of data during workflow execution and the number of iterations of the data placement algorithm in finding the optimal placement location.

\begin{figure}[!t]
	\centering
	\includegraphics[width=8 cm]{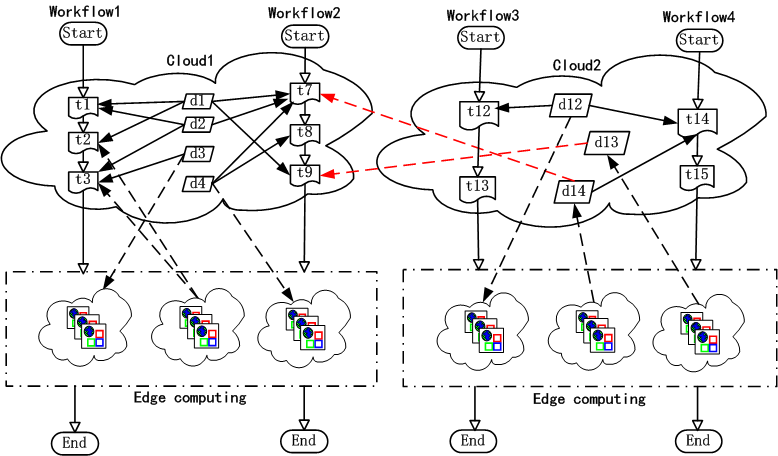}
	\caption{{Sample of data placement for scientific workflow in heterogeneous edge-cloud computing environments.}}
\end{figure}

\begin{table}[!t]
	\centering
	\caption{\\Dataset Size of Data-Sharing Scientific Workflows}
	\label{tab:performance_comparison}
	\begin{tabular}{p{1.1cm}p{0.2cm}p{0.2cm}p{0.2cm}p{0.2cm}p{0.2cm}p{0.2cm}p{0.2cm}p{0.2cm}p{0.2cm}p{0.2cm}p{0.2cm}}	
		\hline
		Dataset& $d_{1}$& $d_{2}$& $d_{3}$& $d_{4}$& $d_{5}$& $d_{6}$& $d_{7}$& $d_{8}$& $d_{9}$& $d_{10}$& $d_{11}$\\
		Size(GB)& 3.1 & 5.4& 2.1& 1.3& 1.1& 2.3& 1.7& 2.1& 1.5& 0.5& 4.0\\
		\hline
	\end{tabular}
\end{table}

\begin{table}[!t]
	\centering
	\caption{\\Placement Cost of Data-Sharing Scientific Workflows}
	\label{tab:performance_comparison}
	\begin{tabular}{p{0.9cm}p{0.275cm}p{0.275cm}p{0.275cm}p{0.275cm}p{0.275cm}p{0.275cm}p{0.275cm}p{0.275cm}p{0.275cm}p{0.275cm}p{0.275cm}}	
		\hline
		Dataset& $d_{1}$& $d_{2}$& $d_{3}$& $d_{4}$& $d_{5}$& $d_{6}$& $d_{7}$& $d_{8}$& $d_{9}$& $d_{10}$& $d_{11}$\\
		Cost(\$)& 0.21 & 0.45 & 0.37 & 0.62 & 0.73 & 0.15 &  0.42 & 1.57 & 1.51& 0.45& 2.05 \\
		\hline
	\end{tabular}
\end{table}

\subsection{Problem Analysis} 
Different data placement policies will directly affect the data transfer time and placement cost. We analyze the data placement problem of scientific workflows which are described in subsection 2.1. In this motivating example, we set the bandwith across the datacenters $\{band_{12},band_{13},band_{23}\}$ as $\{10 M/s, 20 M/s, 150 M/s\}$~\cite{meng2019dedas}. Also, for simplicity, we assume that placement costs of these datasets as table 2. As shown in Figure 3 and Figure 5, there are four data placement strategy lead to four different results. The result of the strategy shown in Figure 3(a) required the datasets to be moved eight times; the amount of data movement was 11.6 GB, the data transfer time was calculated to be 600 s, and data placement cost was 130 dollors. The strategy described in Figure 3(b) was found to have six data movements; the amount of data movement was 8.4 GB, the data transfer time was approximately 280 s, and data placement cost was 60 dollars. the strategy used in Figure 5(a) required the datasets to moved six times; the amount of data movement was 8.4 GB, the data transfer time was approximately 380 s, and data placement cost was 60 dollars. Furthermore, the strategy described in Figure 5(b) was found to have six data movements; the amount of data movement was 8.4 GB, the data transfer time was approximately 280 s, and data placement cost was 60 dollars. Compared with the data placement results of these four strategies, we can analyze that other results achieved through data placement strategy can be better than Fig 3(a). If the data transfer time is chosen as the index, the data placement strategy adopted in Figure 2(b) gets the best result. Meanwhile, if the strategy adopted takes the cost of data placement as the index, the data placement strategy adopted in Figure 2(c) is undoubtedly the best. However, the tradeoff between data transfer time and data placement cost is often considered in the actual data placement process, and Figure 4 is probably the best strategy, although neither transmission time nor data placement cost is minimal. Table 3 shows the final data placement map of datasets under the motivating example in these four strategies.
\begin{figure}[!t]
	\centering
	\includegraphics[width=8 cm]{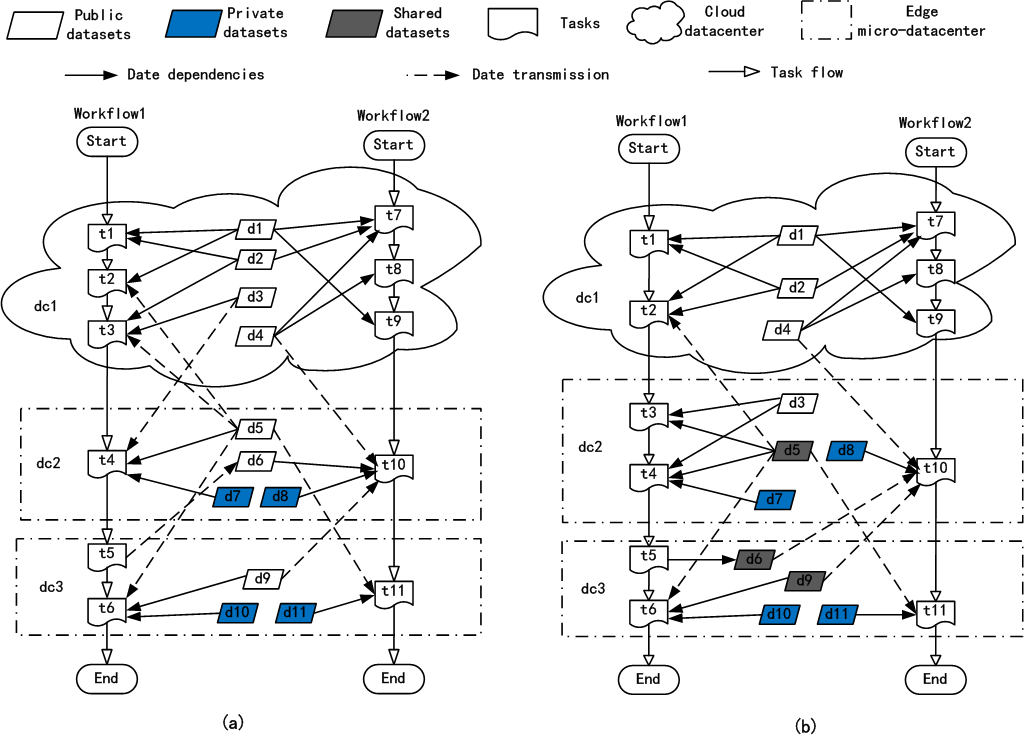}
	\caption{Sample of data placement by two strategies.}
\end{figure} 
\begin{table}[!t]
	\centering
	\caption{\\The Final Placement Location of Each Dataset}
	\label{tab:performance_comparison}
	\begin{tabular}{p{1.4cm}p{0.2cm}p{0.2cm}p{0.2cm}p{0.2cm}p{0.2cm}p{0.2cm}p{0.2cm}p{0.2cm}p{0.2cm}p{0.2cm}p{0.2cm}}	
		\hline
		Dataset& $d_{1}$& $d_{2}$& $d_{3}$& $d_{4}$& $d_{5}$& $d_{6}$& $d_{7}$& $d_{8}$& $d_{9}$& $d_{10}$& $d_{11}$\\
		\hline
		$DC_{fig3(a)}$& $dc_{1}$& $dc_{1}$& $dc_{1}$& $dc_{1}$& $dc_{2}$& $dc_{2}$& $dc_{2}$& $dc_{2}$& $dc_{3}$& $dc_{3}$& $dc_{3}$\\
		$DC_{fig3(b)}$& $dc_{1}$& $dc_{1}$& $dc_{2}$& $dc_{1}$& $dc_{2}$& $dc_{3}$& $dc_{2}$& $dc_{2}$& $dc_{3}$& $dc_{3}$& $dc_{3}$\\
		$DC_{fig5(a)}$& $dc_{1}$& $dc_{1}$& $dc_{1}$& $dc_{1}$& $dc_{2}$& $dc_{2}$& $dc_{2}$& $dc_{2}$& $dc_{3}$& $dc_{3}$& $dc_{3}$\\
		$DC_{fig5(b)}$& $dc_{1}$& $dc_{1}$& $dc_{2}$& $dc_{1}$& $dc_{2}$& $dc_{3}$& $dc_{2}$& $dc_{2}$& $dc_{3}$& $dc_{3}$& $dc_{3}$\\
		\hline
	\end{tabular}
\end{table}

The reason of these four different data placement results are from the heterogeneous edge-cloud computing environments and heterogeneity of datasets in multi-region scenario. During the execution of multiple workflows, we conducted a data placement model, which considers multiple cloud datacenters and multiple scientific workflows. while proposing a data placement strategy based on DE-DPSO, which places datasets while considering not only the bandwidth between datacenters, the number of edge micro-datacenters, and the storage capacity of edge micro-datacenters but also the sharing of data during workflow execution and the number of iterations of the data placement algorithm in finding the optimal placement location.

\section{FORMOLATION FOR DATA PLACEMENT}
In this section, we give fundamental definitions of data-sharing scientific workflows in the heterogeneous edge-cloud computing environments to construct a data placement model for the proposed data placement strategy. It is worth noting that the core purpose of this strategy for all scientific workflows in this model is to optimize the  placement costs for datasets and also to minimize the data transfer time during workflows execution while satisfying the speed of finding the final location for datasets and intrinsic properties of each datacenter (such as storage or computing resource).

\subsection{Scientific  Workflow Definition}
 
Generally, scientific workflows contain numerous tasks linked through the dependency relationship of datasets. Its can be described as data-intensive parallel services and applications.  

{\bf {Definition 1.}} Scientific workflow.
In the study, we describe scientific workflows $W$ as $\{W_{1}, W_{2},..,W_{l}\}$, where $l$ is the number of scientific workflows. Each scientific workflow can be described as a directed acyclic graph $W_{k} = (T,R,DS)$, where $T=\{t_{1},t_{2},...,t_{r}\}$ represents the task set in $W_{k}$, which contains $r$ tasks. An adjacency matrix $R$ represents the relationship between tasks in the task set $T$, and $R_{i,j} = 0$ denotes that task $t_{i}$ has no relationship with task $t_{j}$, $R_{i,j} = 1$ denotes that task $t_{i}$ precedes another task  $t_{j}$, and $DS = \{d_{1},d_{2},...,d_{n}\}$ denotes all datasets in these scientific workflows.

{\bf {Definition 2.}} Datasets. While the state-of-the-art model only requires a scientific workflow, all datasets in the heterogeneous edge-cloud computing environment fall into two categories:  public datasets and private datasets. Due to the confidentiality of scientific datasets  or the particularity of some datasets, private datasets cannot be flexibly transferred and allocated from different datacenters. In other word, they must be stored in particular  micro-datacenters according to their size, ownership or management needs. On the contrary, different research cooperation institutions can flexibly distribute and share by transferring these datasets from any datacenters during the execution of a scientific workflow in this environment. Hence, a dataset $d_{i}$ can be described as $<dsize_{i}, cn_{i}, dc_{i}, pf_{i}>$, where $dsize_{i}$ represents the size of $d_{i}$, and $cn_{i}$ represents the task set that either generates the dataset $d_{i}$ or needs to be generated from this dataset; $d_{i}.dc_{i}$ is the datacenter that stores the dataset $d_{i}$, and $pf$ is a flag that indicates whether or not $d_{i}$ is a private dataset, where $pf = 0$ denotes a public dataset and $pf = 1$ denotes a private dataset.

{\bf {Definition 3.}} Task.  In a scientific workflow $W_{k}$, the task can also be falled into two categories: input datasets and output datasets. While a task is being executed, the datasets associated with it must have a corresponding location in the model.
The task $t_{i}$ is described as $<iDS_{i},oDS_{i}, dc_{i}>$, where $iDS_{i}$ denotes the input datasets of $t_{i}$, $oDS_{i}$ denotes the collection of output datasets, and $t_{i}.dc_{i}$ represents the placement location at which datacenter the task was scheduled. There is a many-to-many map between the task set and data set; in simple terms, an item of data can be used by multiple tasks, and conversely, a task can also request or generate multiple datasets. It is worth noting that because private datasets can only exist in a particular edge micro-datacenter, the tasks which require or generate private datasets are usually separated out for the execution of different workflows. In the execution of a scientific workflow, a task can only be executed if it possesses all the datasets it requires.


{\bf {Definition 4.}} Data placement map. We define $MS=(W,D,DC,Map)$ as the data placement map, where $Map$ can be described as dataset--datacenter mapping. For a scientific workflow $W_{k}$, the $Map_{W_{k}}$ can be divided into two cases, depending on the datasets. For a private data placement map, which can be formularized as $Map.pri = \bigcup_{d_{i} \in {d_{i}.pri}}\{d_{i} \to d_{i}.dc \}$, these datasets can be mapped directly to their fixed locations (i.e.,edge micro-datacenters). For a public data placement map $Map.pub = \bigcup_{d_{i} \in {d_{i}.pub}}\{d_{i} \to d_{i}.dc \}$, these datasets can be mapped by data placement algorithms. 

It is worth noting that because task transfer time and placement cost compared to data transfer time and placement cost, a task is a piece of code which is normally very small so as to be omitted. So we assume that there is no additional task transfer time and placement cost when the task is placed to datacenters. 

\subsection{Data Placement Model}

In order to improve the efficiency of data placement in the model, minimizing the data transfer time and placement cost, we propose a data placement model which considers data-sharing scientific workflows. The placement model takes into account shared datasets not only both within individual region and among multiple regions, but also a workflows and among multiple workflows. The same as private datasets, the shared datasets also play an important role in the placement for our model in the whole system. Hence, when we construct the data-shared placement model,  the definitions of scientific workflows need to be redefined. For example, when we define the dataset in Definition 2, only one attribute $pf$ is not express whether the data is shared or not. 

The data-shared placement model is constructed to find good data placement solutions, which not from just individual scientific workflow or a  geographical region only. By placing more shared datasets together, we can get better to reduce data transfer time and economize data placement cost in heterogeneous edge-cloud computing environments. 

{\bf {Definition 5.}} Datasets. While the execution of scientific workflows in the edge computing, the previous data placement model only requires a scientific workflow. The proposed model not only does the data need to be divided into public and private datasets, but also need to be made between datasets that are shared across multiple workflows. Futhermore, the model can find better data placement solutions from  multiple geo-distributed data-sharing scientific workflows. Hence, we describe a dataset $d_{i}$ as $<dsize_{i}, cn_{i}, dc_{i}, pf_{i}, sf_{i}, lg_{i}>$, where $dsize_{i}$ represents the size of $d_{i}$, and $cn_{i}$ represents the task set that either generates the dataset $d_{i}$ or needs to be generated from this dataset; $d_{i}.dc_{i}$ is the datacenter that stores the dataset $d_{i}$, and $pf$ is a flag that indicates whether or not $d_{i}$ is a private dataset, where $pf = 0$ denotes a public dataset and $pf = 1$ denotes a private dataset.  $sf_{i}$ indicates whether or not the dataset  $d_{i}$ is shared between different workflows, where $sf_{i} = 0$ denotes an unshared dataset and $sf_{i} = 1$ represents a shared dataset. The last attribute, $lg_{i}$, denotes whether or not the dataset $d_{i}$ is shared between different geo-region workflows, where $lg_{i} = 1$ denotes the dataset need be shared in different geo-region workflows during 
the cooperation of different scientific institutions.

{\bf {Definition 6.}} Task.  In the data-sharing placement model, we denote task set $T$  in the whole system as $<iDS,oDS, dc>$, where $iDS$ denotes the input datasets, $oDS$ denotes the collection of output datasets, and $dc$ represents the placement location.  Different from the tranditional defination of task, for each workflow $W_{k}$, a task $t_{i}$ should be described more detail. We replace $iDS_{i}$ as $t_{i}.iDS_{w}$, and replace $oDS_{i}$ as $t_{i}.oDS_{w}$. Accourding to whether the task requires scheduling shared datasets, the  $t_{i}.iDS_{w}$ is denoted as $t_{i}.iDS_{w}.sf_{1}$ and $t_{i}.iDS_{w}.sf_{0}$. Meanwhile, if the task requires sheduling private datasets, it is denoted as $t_{i}.iDS_{w}.pf_{1}$; otherwise, it is represented as 
$t_{i}.iDS_{w}.pf_{1}$. Futhermore,  we describe $t_{i}.iDS_{w}.lg_{1}$ as the task requires scheduling shared datasets across different regions;  $t_{i}.iDS_{w}.lg_{1}$ is described as their datasets from a same region.

{\bf {Definition 7.}} Data Placement. In this model,  $S=(W,D,DC,Map,T_{trans},C_{cost},N_{iter})$ defines as the data placement, where $Map$ can be described as dataset--datacenter mapping. The $Map$ = $ <pri,pub>$ denotes global data placement map, which is similary with Defination 4. For a private data placement map, which can be formularized as $Map.pri = \bigcup_{d_{i} \in {d_{i}.pri}}\{d_{i} \to d_{i}.dc \}$, these datasets can be mapped directly to their fixed locations (i.e.,edge micro-datacenters). For a public data placement map $Map.pub = \bigcup_{d_{i} \in {d_{i}.pub}}\{d_{i} \to d_{i}.dc \}$, these datasets can be mapped by data placement algorithms. Specially, we futher represent $Map.pub$ as $<Map.pub.sh>$ and  $<Map.pub,ush>$, where $Map.pub.sh = \bigcup_{d_{i} \in {d_{i}.pub}}\{d_{i} \to d_{i}.dc | sf_{i} = 1\}$ and $Map.pub.ush = \bigcup_{d_{i} \in {d_{i}.pub}}\{d_{i} \to d_{i}.dc | sf_{i} = 0\}$. $<Map.pub.sh>$ can be detailedly divided  $<Map.pub.sh.lg>$ and  $<Map.pub.sh.lg>$, where $Map.pub.sh.lg = \bigcup_{d_{i} \in {d_{i}.pub}}\{d_{i} \to d_{i}.dc | sf_{i} = 1 \& lg_{i} = 1\}$ and $Map.pub.sh.ulg = \bigcup_{d_{i} \in {d_{i}.pub}}\{d_{i} \to d_{i}.dc | sf_{i} = 1 \& lg_{i} = 0\}$.

During the execution of all the workflows in the model, $T_{trans}$ denotes the total data transfer time and $C_{cost}$ represents the map of data placement cost. We use formula (1) to calculate the data thansfer time $T_{trans}$ ,  $C_{cost}$ can be calculated by (2), and $N_{iter}$ can be calculated by iteration times of data placement algorithm (DPA). 
\begin{equation}
T_{trans} = \sum_{i=1}^{\left|DC\right|}\sum_{j\neq i}^{\left|DC\right|}\sum_{k=1}^{\left|D\right|}\dfrac{dsize_{k}}{band_{ij}} \cdot g_{ijk}
\end{equation}
where  $g_{ijk}$ is used to discern whether a dataset $d_{k}$ is being transferred from different datacenters throughout the process of data scheduling; $g_{ijk} = 0$ indicates that $d_{k}$ is always in the same datacenter.

\begin{equation}
\begin{cases}
C_{cost} = Cost_{wl}^{rt}(Map.pub) + Cost_{wl}^{rt}(Map.pri)
\\
\vec Cost_{wl}^{rt}(Map.pub)=
\begin{cases}

\begin{aligned}
\vec Cost_{wl}^{rt}(Map.pub.sh)
\end{aligned}
\\
\vec  Cost_{wl}^{rt}(Map.pub.ush)
\end{cases}
\\
\vec Cost_{wl}^{rt}(Map.pub.sh) =
\begin{cases}

\begin{aligned}
\vec Cost_{wl}^{rt}(Map.pub.sh.lg)
\end{aligned}
\\
\vec Cost_{wl}^{rt}(Map.pub.sh.ulg)
\end{cases}
\end{cases}
\end{equation}

where  $g_{ijk}$ is used to discern whether a dataset $d_{k}$ is being transferred from different datacenters throughout the process of data scheduling; $g_{ijk} = 0$ indicates that $d_{k}$ is always in the same datacenter. where  $g_{ijk}$ is used to discern whether a dataset $d_{k}$ is being transferred from different datacenters throughout the process of data scheduling; $g_{ijk} = 0$ indicates that $d_{k}$ is always in the same datacenter.

Using the shortest map time to obtaining the lowest data transfer time and placement cost  is the optimal workflow strategy for data placement mapping. Hence, the characteristics of data and the data placement strategy are the two most important factors influencing our model. We have considered the privacy and sharing of the data, and we will describe our data placement strategy in the next section. Therefore, our data placement model comprehensively calculates the data transfer time and data placement cost for scientific workflows in heterogeneous edge-cloud computing environments.

\section{Data Placement Strategy}

A model has been constructed based on real-life parameters, which involve  multiple geo-distributed data-sharing scientific workflows. In this section, a novel data placement strategy based on the model is described, which provides the algorithm for finding a better data placement map, which can maximize the speed of finding the optimal location, to minimize data transfer time and data placement cost. We describe our data placement strategie as two stages: first, a data placement algorithm(DE-DPSO-DPA) is designed to determine the final locations of public datasets. Afterwards, the proposed data placement strategy (DE-DPSO-DPS)  specifically based on DE-DPSO-DPA is described.
\subsection{DE-DPSO-DPA}
The PSO algorithm and DE algorithm are both heuristic search algorithms based on groups, proposed by \cite{kennedy1995particle} and \cite{qin2008differential}, respectively. While elaborating on these algorithms, which are beyond the scope of this paper, our DE-DPSO algorithm does not only enables traditional PSO to solve discrete problems like data placement through problem coding, but also skillfully combines an efficient global optimization DE algorithm.
\begin{algorithm}[!t]
	\caption{DE-DPSO-DPA}
	\label{alg:A}
	\begin{algorithmic}[1]
		\REQUIRE~~\\  T(itermax) , t(current iteration),	n (particles), \\ D (the dimension) , F(scaling factor), $Cr_{g}$, $Cr_{p}$
		
		\ENSURE~~\\ $Res$ (the best optimal solution)
		\STATE Set parameters and Initialize all datasets' placement;
		\STATE 	Set particle dimension $H=|D.pub|$;
		{  \STATE \bf	for}  i : 1 to swarm size n {\bf do}
		{  \STATE \hspace*{0.1in} \bf	for}  d : 1 to $H$ {\bf do}
		\STATE 	\hspace*{0.2in} Initialize $x_{id}^{k}$ randomly;
		{  \STATE \hspace*{0.1in} \bf End for}\\
		\STATE \hspace*{0.1in}	Initialize pbest\\
		{  \STATE  \bf End for}\\
		\STATE 	Initialize gbest\\
		\WHILE{t $<=$ T }
		
		{  \STATE \bf	for}  i = 1 to n {\bf do}
		{\STATE \hspace*{0.1in} Select a, b randomly from particles and a $\neq$ b; }
		{\STATE \hspace*{0.1in} mutation($ x_{i,t-1}, F, x_{a,t-1}, x_{b,t-1}$) by Equation (3)}
		{\STATE \hspace*{0.1in} crossover($x_{pbest,t-1}$,$u_{i,t}$,$Cr_{p}$) by Equation (4)} 
		{\STATE \hspace*{0.1in} crossover($x_{gbest,t-1}$,$u_{i,t}$,$Cr_{g}$) by Equation (4)} 
		{\STATE \hspace*{0.1in} Select $x_{i,t}$ by Equation (5)  }
		{\STATE \bf  end for}	
		{\STATE  t= t+1}		
		\ENDWHILE
		{ \STATE Update the best optimal solution Res} 
		{\STATE  \bf  Output Result}
	\end{algorithmic}
\end{algorithm}
\subsubsection{Problem Encoding} 
Inspired by the coding strategy mentioned in~\cite{lin2019time,8634658}, a discrete coding strategy is provided for the data placement problem to satisfy the well-known principles of completeness, non-redundancy, and viability. Most meta-heuristic algorithms filter the optimal solution by generating n-dimensional candidate particles.
In our data placement problem, a particle maps a solution for data-sharing scientific workflows in the proposed distributed model. The $i$th particle in the algorithm's $t$th iteration is formulated as
\begin{equation}
X_{i,t} = \{x_{i,t}^{1},x_{i,t}^{2},...,x_{i,t}^{d}\}
\end{equation}
where $d$ is the dimension of this particle and represents the number of datasets in the model. Meanwhile, $x_{i,t}^{k}$ denotes the placement location of the $k$th dataset after the $t$th iteration of the algorithm. It is worth noting that, for a particle in this model, there are $Q$ dimensions representing the private datasets are fixed, and the $H$ dimensions representing datasets are shared for multiple workflows.

After determining the correspondence between each particle and the candidate solution in our model, a data placement algorithm, which finding the optimal location in the shortest time, is proposed such that the exposed dataset could be placed in the appropriate datacenter, in order to achieve the goal of minimizing data transmission time and data placement cost. 
\subsubsection{Algorithm  description}
The traditional PSO algorithm is a classical heuristic algorithm, and its update strategy is based on the velocity and position of the particles. When it is used as a data placement algorithm, there are two problems: 
\begin{itemize}
	\item It is easy to fall into local optimization.
	\item It cannot handle discrete problems, such as the one in this study.
\end{itemize}

To address these issues, the DE algorithm is used to expand its search capability, and the algorithm is also discretized based on the application scenarios. The pseudocode of our DE-DPSO-DPA algorithm for scientific workflows is presented in Algorithm 1. Specifically, at the first part (lines 1-9), all the datasets and the algorithm parameters are initialized. The steps are the preprogressing of placing shared datasets, and the algorithm evaluates the current fitness of particles and also updates best values with the best result. Thereafter, according to the steps described in 10)$\sim$17), the update strategy for each particle at every iteration adapts the mutation and crossover operators of the DE to update the best global solution, which effectively prevents the algorithm from falling into the local optimal solution. The mutation strategy for the $i$th particle at the $t$th iteration is formulated as follows.
\begin{equation}
u_{i,t} = x_{i,t-1} \oplus F \odot (x_{a,t-1} \ominus x_{b,t-1})
\end{equation}
where $u_{i,t}$ is a new feasible particle and $F$ is the scale factor. After the mutation, the generated particle is changed by a crossover strategy. For the individual cognition and social cognition components, the crossover operator of our algorithm is shown in Equation (4).
\begin{equation}
\begin{cases}
\vec y = crossover(\vec x_{1}, \vec x_{2},prob)
\\
\vec y[i] =
\begin{cases}

\begin{aligned}
\vec x_{1}[i] \qquad if \quad Random.r < prob
\end{aligned}
\\
\vec x_{2}[i] \qquad \ if \quad Random.r \geq prob
\end{cases}
\end{cases}
\end{equation}
where  $Random.r$ is a random factor between 0 and 1. $prob$ is a parameter used to control the extent of crossover operations. In order to get the optimal solution, the operation is executed twice in each algorithm's iteration. Specifically, $Cr_{p}$ denotes the crossover parameter, which controls the distance between the current particle position and the local optimal position. Meanwhile, $Cr_{g}$ is also a parameter that proportionally selects indexes in an old particle and replaces the segment between them with the $gbest$ particle segment. $\vec y$, $\vec x_{1}$ and $\vec x_{2}$ also represent particles in different cases. The final particle, $\vec y$, obtained by the above operations at the $t$th iteration is denoted as $w_{i,t}$.

The optimality of a particle is usually measured by the fitness function: the smaller it is, the better the performance. In the present data placement problem, clearly, the better particle must have a smaller data transfer time and a higher speed of finding the final location of all the datasets. In the discrete encoding method, the fitness value $fit_{1}()$ and $fit_{2}()$ are defined as the data transmission time and placement cost respectively.
In summary, the update strategy can be described as follows:
\begin{equation}
x_{i,t} =
\begin{cases}
\begin{aligned}
w_{i,t} \qquad if \quad fit(w_{i,t}) < fit(gbest)
\end{aligned} 
\\
x_{i,t-1}\qquad if\quad fit(w_{i,t}) \geq fit(gbest)
\end{cases}
\end{equation}

\begin{equation}
x_{i,t} =
\begin{cases}
\begin{aligned}
w_{i,t} \qquad if \quad fit(w_{i,t}) < fit(gbest)
\end{aligned} 
\\
x_{i,t-1}\qquad if\quad fit(w_{i,t}) \geq fit(gbest)
\end{cases}
\end{equation}

where the $T_{trans(X_i)}$ represents the total transfer time required for all data during the allocation of the datacenter, and the $N_{iter}$ represents the number of iterations when the algorithm converges and can be used to quantify the speed of the strategy in finding the optimal position (i.e. the fewer iterations, the faster the speed).

\begin{algorithm}[!t]
	\caption{DE-DPSO-DPS}
	\label{alg:A}
	\begin{algorithmic}[1]
		\REQUIRE~~\\  initial datasets $DS$, Tasks $T$,  Datacenters $DC$
		\ENSURE~~\\ $PM$ (data placement map), Data transfer time $T_{trans}$, Number of iterations $N_{iter}$	
		\STATE Initial Data transfer time $T_{trans}$ = 0, Data placement cost  $C_{cost}$ = 0, number of algorithm's iteration $N_{iter}$ = 0,
		\STATE Divide datasets $DS$ into $DS.pub$ and $DS.pri$;
		\STATE Allocate datasets $DS.pri$ to $DC$;
		\STATE Divide datasets $DS.pub$ into $DS.ush$ and  $DS.sh$;
		\STATE allocate datasets $D.pub$ into $DC$;
		\STATE DE-DPSO-DPA;\\	
		\STATE Calculate $T_{trans}$, $C_{cost}$ and $N_{iter}$;
		\STATE \bf{Output Result}
	\end{algorithmic}
\end{algorithm}
\subsection{DE-DPSO-DPS}
Algorithm 2 is designed to describe our data placement strategy for data-sharing scientific workflows. It consists of three parts: data placement of private datasets (lines 1-3), data placement of public  datasets(lines 4-6), and the combination of the former two parts (line 7-8).

In part one, after initializing the current storage of all the datacenters, the total data transfer time, data placement cost and the number of DE-DPSO-DPA iterations to 0, all datasets are classified into public datasets and private datasets. Subsequently, these private datasets are distributed to fixed datacenters. In part two, the remaining public datasets are divided into unshared and shared datasets. These unshared and shared datasets are then respectively distributed to proper datacenters according to the DE-DPSO-DPA. In part three, this algorithm assembles the maps, the total data transmission time, and the number of iterations of all the datasets. 
There have been two improvements: (1) all data placement solutions are transformed into a closed annular searching space in order to ensure that they are effective and are mapped to actual datacenters; (2) the particle velocity speed is constrained based on the dataset characteristics (e.g., public/private and shared/unshared) in order to guarantee that DE-DPSO-DPS can locate appropriate data placement solutions.%

The problem with data placement strategies for data-sharing scientific workflows in heterogeneous edge-cloud computing environments can be represented by Equation(6).
It's core purpose is to pursue a minimum total data transmission time and placement cost while satisfying the storage capacity constraint for each datacenter. 
\begin{equation}
\begin{cases}
\begin{aligned}
Minimize \ T_{trans}
\end{aligned}
\\
\begin{aligned}
Minimize \ Cost_{trans}
\end{aligned}
\\
subject \ to \  \forall i, \sum_{j=1}^{\left|D\right|} d_{j} \cdot l_{ij} \leq capacity_{i}
\end{cases}
\end{equation}
where $l_{ij}$ is a flag which represents whether the datacenter $dc_{i}$ stores the dataset $d_{j}$; $l_{ij} = 1$ indicates that it does, while $l_{ij} = 0$ indicates that it does not. $capacity_{i}$ denotes the storage capacity of the $i$th datacenter. 

\begin{table*}[!t]
	\begin{threeparttable}
		\centering
		\fontsize{10}{18}\selectfont
		\caption{Comparison of Different Strategies with Different Bandwidths}
		\label{tab:performance_comparison}
		\begin{tabular}{p{2.5cm}cc|cc|cc|cc|cc}
			
			\hline
			
			\multicolumn{11}{c}{\bf{\qquad\qquad\qquad Bandwidth of Edge micro-Datacenters}}\cr\cline{2-11}

			\multirow{1}{*}{\bf{Algorithms}}&
			\multicolumn{2}{c}{\bf{0.5}}&
			\multicolumn{2}{c}{\bf{0.8}}&
			\multicolumn{2}{c}{\bf{1.5}}&
			\multicolumn{2}{c}{\bf{3}}&
			\multicolumn{2}{c}{\bf{5}}\cr\cline{2-11}
			&$T_{trans}$&$N_{iter}$&$T_{trans}$&$N_{iter}$&$T_{trans}$&$N_{iter}$&$T_{trans}$&$N_{iter}$
			&$T_{trans}$&$N_{iter}$
			\cr
			\hline
			Random
			&36564.17
			&  -
			&26032.87
			&  -
			&14787.45
			&  -
			&9116.71
			&  -
			&6788.63
			
			&-\cr\hline
			DE-DPS
			&7883.97
			&1707
			&6182.27
			&1442
			&4845.19
			&1186
			&4088.46
			&1042
			&3787.79
			
			&1489\cr\hline
			DPSO-DPS
			&8092.25
			&165
			&6223.92
			&182
			&4885.87
			&135
			&4114.91
			&117
			&3812.82
			
			&194\cr\hline
			GA-DPSO-DPS
			&7827.42
			&1315
			&6112.22
			&1384
			&4830.79
			&1664
			&4081.86
			&1523
			&3786.71
			&1410
			
			\cr\hline
			{\bf DE-DPSO-DPS}&{\bf 7818.04}&{\bf 211}&{\bf 6069.18}&{\bf 351}&{\bf 4830.34}&{\bf 327}&{\bf 4082.89}&{\bf 265}&{\bf 3785.50}&{\bf 198}\cr\hline
			
		\end{tabular}
		\begin{tablenotes}
			\footnotesize
			\item[*] The unit of $T_{trans}$ is seconds(s), and $N_{iter}$ can be calculated by iteration times of data placement algorithm (DPA).
		\end{tablenotes}
	\end{threeparttable}
\end{table*}
\begin{table*}[!t]
	\centering
	\fontsize{10}{18}\selectfont
	\caption{Comparison of Different Strategies with Different Storage Capacity of Edge micro-Datacenters}
	\label{tab:performance_comparison}
	\begin{tabular}{p{2.5cm}cc|cc|cc|cc|cc}
		
		\hline
		
		\multicolumn{11}{c}{\bf{\qquad\qquad\qquad Storage Capacity of Edge micro-Datacenters}}\cr\cline{2-11}

		\multirow{1}{*}{\bf{Algorithms}}&
		\multicolumn{2}{c}{\bf{150G}}&
		\multicolumn{2}{c}{\bf{200G}}&
		\multicolumn{2}{c}{\bf{250G}}&
		\multicolumn{2}{c}{\bf{300G}}&
		\multicolumn{2}{c}{\bf{350G}}\cr\cline{2-11}
		&$T_{trans}$&$N_{iter}$&$T_{trans}$&$N_{iter}$&$T_{trans}$&$N_{iter}$&$T_{trans}$&$N_{iter}$
		&$T_{trans}$&$N_{iter}$
		
		\cr
		\hline
		Random
		&19267.53
		&  -
		&13947.78
		&  -
		&14300.76
		&  -
		&13413.87
		&  -
		&12729.17
		
		&-\cr\hline
		DE-DPS
		&5610.33
		&968
		&5483.35
		&1038
		&4460.15
		&1272
		&4383.37
		&1467
		&4358.24
		
		&769\cr\hline
		DPSO-DPS
		&5685.15
		&151
		&5428.95
		&105
		&5877.73
		&109
		&4315.18
		&137
		&4268.19
		
		&105\cr\hline
		GA-DPSO-DPS
		&5601.00
		&1485
		&5385.59
		&1263
		&4479.62
		&1549
		&4258.96
		&1623
		&4164.98
		&1637
		
		\cr\hline
		{\bf DE-DPSO-DPS}&{\bf 5585.15}&{\bf 218}&{\bf 5382.84}&{\bf 323}&{\bf 4466.68}&{\bf 312}&{\bf 4251.91}&{\bf 251}&{\bf 4161.89}&{\bf 225}\cr\hline
		
	\end{tabular}
\end{table*}

\section{Experimental results and Analysis}

In this section, several simulation experiments, designed to evaluate the effectiveness of the proposed data placement strategy, are described and the impact factors in our data placement model are discussed. Comparing the results with those from other strategies and also considering the different scenarios of these factors, the advantages of our proposed strategy in this context, as well as the impact factors, are evaluated. All the experiments were conducted on a machine with the following specifications: an Intel(R) Core(TM) i7--4790 CPU @ 3.40 GHz, 16 GB of RAM, Windows10(64bit), and IntelliJ IDEA2019.2.4.
\subsection{Experimental Setup}
In the developed DE-DPSO-DPA,  the initial
population size was set to 100, the maximum number of iterations was set to 2000, the scaling factor was set to 0.15, and $Cr_{g}$ and  $Cr_{p}$ were set to 0.1 and 0.1, respectively. The synthetic workflows from Montage in astronomy released by Bharathi et al~\cite{bharathi2008characterization} were used, and both the number of datasets and the structures differed in them.  In order to discuss the impact factors in our model, the number of edge micro-datacenters was changed from three to five, and the number of cloud datacenters was set at two. The storage capacity of edge micro-datacenters was the same in the same simulation, varying from 150 GB to 300 GB in steps of 50 GB. In addition, the bandwidth between different datacenters in basic experiment was set as follows(its unit is M/s): 
\begin{equation}
Bandwidth = {
	\left[ \begin{array}{ccccc}
	\sim & 5  & 5  & 5   & 5  \\
	5 & \sim  & 20 &20   &20   \\
	5 & 20 & \sim  &100  &150 \\
	5 & 20 &100 &  \sim &200   \\
	5 & 20 &150 &200  & \sim   \\	
	\end{array} 
	\right ]}
\end{equation}

the bandwidth between two cloud datacenters was set as 5 M/s, the bandwidth between an edge micro-datacenter and a cloud datacenter was 20 M/s, and the bandwidth between different edge micro-datacenters were set as \{100 M/s, 150 M/s, 200 M/s\}. In the basic experiment, there are three edge micro-datacenters and their storage capacity was set as 150GB. To compare the performance of different bandwidths across different edge micro-datacenters of these data placement strategies, we observed the bandwidth across different edge micro-datacenters is \{0.5, 0.8, 1.5, 3, 5\} times faster than the bandwidth in the basic experiment. Futhermore, the general workflows are generated by the given workflow settings listed in Table 7. For each examination, both the number of specific and general workflows in the sample space is 100. And the average values of executing all workflows are shown in Figs. 4, 5, 6, 7, and 8.

\subsection{Performance Comparison}
Four other data placement strategies were compared with our DE-DPSO-DPS. The deployment of data in the other strategies was based on the following algorithms: random, DE, DPSO, and GA-DPSO. In view of the fact that, there are several similarities between the distributed cloud computing environment and our heterogeneous edge-cloud computing environment, the methods used in the former can be refined and applied to our model and compared to our strategy. When the strategy is random, the number of iterations is 1, so the $N_{iter}$ was ignored. DE, DPSO, and GA-DPSO are meta-heuristic algorithms, which need to converge and may generate different results in each experiment. According to~\cite{lin2019time}, if the fitness values are consistent for over 80 consecutive iterations, the algorithm is considered to have converged. The data transfer time and the speed of finding the final location for all datasets are measured as the average of 100 repeated experiments.

\subsubsection{Variation in Bandwidth and Storage Capacity of Edge micro-Datacenters}
Table III and Table IV show the results of comparing the different bandwidths and different storage capacities, of the edge computing datacenters for the five data placement strategies in achieving the optimal result for our scientific workflows. As the bandwidth across different datacenters increased, the number of their DPA's iteration times did not fluctuate very much, and the data transmission time decreased significantly. As shown in table IV, with the increase of the storage capacities, both the data transmission time and the number of iterations are reduced. For data transmission time, no matter how the impact factors change, the DE-DPSO-DPS is the most optimal. For the iteration times of their DPA, only DPSO-DPS is slightly less than DE-DPSO-DPS, which is determined by the fact that DPSO is extremely prone to local optimization.

\subsubsection{Variation in Ratio of Fixed and Shared Datasets}
Table III and Table IV show the results of comparing the different bandwidths and different storage capacities, of the edge computing datacenters for the five data placement strategies in achieving the optimal result for our scientific workflows. As the bandwidth across different datacenters increased, the number of their DPA's iteration times did not fluctuate very much, and the data transmission time decreased significantly. As shown in table IV, with the increase of the storage capacities, both the data transmission time and the number of iterations are reduced. For data transmission time, no matter how the impact factors change, the DE-DPSO-DPS is the most optimal. For the iteration times of their DPA, only DPSO-DPS is slightly less than DE-DPSO-DPS, which is determined by the fact that DPSO is extremely prone to local optimization.

\subsubsection{Variation in Number of Workflow}
It can be observed that the growth of data transfer cost
obtained from these three data placement strategies is stable with the increase of the number of data-sharing scientific cloud workflows. And the performance of the workflow- level strategy is better than that of random and task-level strategies. In fact, our strategy reduces the total data transfer cost by 16.70 and 10.67 percent on average respectively in comparison to the random and task-level strategies. But for the task-level data placement strategy, it only cuts the cost down by 6.77 percent on average in comparison to the ran- dom strategy. In conclusion, for general data-sharing scientific cloud workflows, the performance of our workflow-level data placement strategy is much better than its task-level and random counterparts. Specifically, the random strategy does not optimize the placement of initial datasets and only distributes generated flexible datasets to the location where they are generated.

\subsubsection{Variation in Number of edge-datacenter}
Figure 4 shows the data transfer time and the number of iterations of different data placement strategies with and without data sharing. Our data-sharing-based model is superior to those in which each workflow is considered separately, in terms of the total data transfer time under the same data placement strategy. Additionally, the proposed data placement strategy results in minimum data transfer time, regardless of whether sharing is considered or not. Futhermore, because all workflows are treated as one, the model that considers shared datasets shows tremendous advantage for their DPA's iteration times in each strategy.


\subsubsection{Variation in Storage Capacity of edge-datacenter}
Figure 6 shows the data transfer time and the number of iterations of different data placement strategies with and without data sharing. Our data-sharing-based model is superior to those in which each workflow is considered separately, in terms of the total data transfer time under the same data placement strategy. Additionally, the proposed data placement strategy results in minimum data transfer time, regardless of whether sharing is considered or not. Futhermore, because all workflows are treated as one, the model that considers shared datasets shows tremendous advantage for their DPA's iteration times in each strategy.

\subsubsection{iteration times of DPAs}
Figure 7 shows the data transfer time and the number of iterations of different data placement strategies with and without data sharing. Our data-sharing-based model is superior to those in which each workflow is considered separately, in terms of the total data transfer time under the same data placement strategy. Additionally, the proposed data placement strategy results in minimum data transfer time, regardless of whether sharing is considered or not. Futhermore, because all workflows are treated as one, the model that considers shared datasets shows tremendous advantage for their DPA's iteration times in each strategy.

\subsubsection{different datasets}
Figure 8 shows the data transfer time and the number of iterations of different data placement strategies with and without data sharing. Our data-sharing-based model is superior to those in which each workflow is considered separately, in terms of the total data transfer time under the same data placement strategy. Additionally, the proposed data placement strategy results in minimum data transfer time, regardless of whether sharing is considered or not. Futhermore, because all workflows are treated as one, the model that considers shared datasets shows tremendous advantage for their DPA's iteration times in each strategy.

\section{Related Work}
Developing a strategy to finding an optimal data placement in a distributed environment has always been a considerable challenge. Whether in traditional storage systems or in the newer cloud and heterogeneous edge-cloud computing systems, the data placement problem has been studied in depth. In this era of highly developed network technology, we require a model which considers real-world scenarios as much as possible. Along with factors, such as large datasets, shared datasets, private datasets stored in fixed edge micro-datacenters, and bandwidth limitations, which have a critical effect on the data transfer time and the speed of finding the final location for data placement.

Like cluster or grid systems, several traditional deployment methods for scientific workflows involve distributed computing environments, which are very expensive and do not satisfy the demands of a practical scenario. Moreover, these deployment environments with low-level resource sharing will lead to significant data transmission delays. In the existing distributed computing systems, researchers have focused on optimizing the simulation models and data transfer time in cloud environments. Nukarapu et al.~\cite{nukarapu2010data} proposed a classic data-intensive scientific workflow system deployed on a distributed platform, which explores the interactions between data placement services and relatively reduced system execution time. Yuan et al.~\cite{yuan2010data} provided a data placement strategy based on k-means and BEA clustering for a scientific workflow that effectively reduced the number of data movements. However, this strategy ignored the difference in the storage capacity of each datacenter. In addition, the number of data movements did not accurately represent the amount of data movement or the actual data transmission status. Wang et al.~\cite{wang2014data} designed a data placement strategy based on k-means clustering for a scientific workflow in cloud environments that considered the data size and dependency. While this approach reduced the number of data movements using a data replication mechanism, it did not formalize the data replication cost.

As per the results of recent studies, scientific cloud workflow systems can satisfy the demands of scientists from different laboratories or regions, and they can also collaborate and carry out their research process more flexibly~\cite{li2019overall}. However, owing to the existence of large-scale dataset interaction and the remote deployment of cloud computing sources, data transmission still requires a significant amount of time during workflow execution. To solve this problem, a time-driven data placement strategy has been developed by combining the advantages of both edge and cloud computing, which also solves the problem of limited resources in edge computing. 

In summary, existing studies have only proposed data placement models and strategies to reduce the data transfer time for individual workflows. In particular, as per the latest discussions on heterogeneous edge-cloud computing environments, these models generally consider only one cloud datacenter and one scientific workflow. However, in the real world, cooperation between scientific organizations across different geographical distributions is common. Therefore, the existence of multiple scientific workflows and multiple cloud data centers should be considered when building a data placement model. In such a model, the data sharing can have a big impact on the placement outcome. Additionally, when formulating a data placement strategy, the data transfer time as well as the speed of data placement must be minimized. In this study, distinguishing from the other existing works, a novel data placement model is constructed according to the environment mentioned above, and a strategy is proposed to minimize data transfer time and optimize the speed of obtaining the final placement of data for data-sharing scientific workflows.

\section{Conclusion}

This study combines advantages of both edge and cloud computing to construct a data placement model, which can balance data transfer time and data placement cost using intelligent computation. A data placement model which considers multiple workflows distributed across different geographic regions is presented, and a data placement algorithm is designed to allocate public datasets. Experimental results express that the data placement strategy based on this algorithm can effectively reduce the data transmission time in the process of scientific workflow operation, as well as increase the speed of finding optimal data placement, as compared to other state-of-the-art algorithms. In addition, through the discussion of the number of edge micro-datacenters; storage capacity of edge micro-datacenters; bandwidth between datacenters; and other factors that may affect the performance of the data placement strategy, the advantages of our strategy, in each case, were analyzed in more detail.

\section*{Acknowledgment}
The work of this paper is supported by National Key Research and Development Program 
of China (2019YFB1405000), National Natural Science Foundation of China under Grant (No. 61873309, No. 61972034,No. 61572137, and No. 61728202), and Shanghai Science and Technology Innovation Action Plan Project under Grant (No.19510710500,No. 18510732000,and No.18510760200).
\bibliographystyle{IEEEtran}
\bibliography{references}

\appendices


\ifCLASSOPTIONcompsoc


\ifCLASSOPTIONcaptionsoff
  \newpage
\fi

\end{document}